\documentclass[nojss]{jss}

\usepackage{amsmath, amssymb, amsfonts, amsthm}
\usepackage{bm}
\usepackage{subfig}
\usepackage{caption}
\usepackage{graphicx}
\usepackage{tikz}
\usetikzlibrary{shapes.misc, positioning}

\author{Berent Ånund Strømnes Lunde\\Equinor\\University of Bergen
   \And Feda Curic\\Equinor
  \And Sondre Sortland\\Equinor

}
\title{\pkg{GraphSPME}: Markov Precision Matrix Estimation 
and Asymptotic Stein-Type Shrinkage}

\Plainauthor{Berent Lunde, Feda Curic, Sondre Sortland}
\Plaintitle{GraphSPME: Markov Precision Matrix Estimation 
and Asymptotic Stein-Type Shrinkage} %% without formatting
\Shorttitle{\pkg{GraphSPME}: Asymptotic Markov Precision} %% a short title (if necessary)

\Abstract{
  \pkg{GraphSPME} is an open source \proglang{Python}, \proglang{R} and \proglang{C++} header-only
  package implementing non-parametric sparse precision matrix estimation along with 
  asymptotic Stein-type shrinkage estimation of the covariance matrix.
  The user defines a potential neighbourhood structure and provides data that potentially are 
  $p>>n$. 
  This paper introduces a novel approach for finding the optimal order (that data allows to estimate)
  of a potential Markov property.
  The algorithm is implemented in the package, alleviating the problem of users making Markov assumptions
  and implementing corresponding complex higher-order neighbourhood structures.
  Estimation is made accurate and stable by simultaneously utilising both Markov properties
  and Stein-type shrinkage.
  Asymptotic results on Stein-type shrinkage ensure that non-singular well conditioned
  matrices are obtained in an automatic manner. 
  Final symmetry conversion creates symmetric positive definite estimates.
  Furthermore, the estimation routine is made efficient and scalable to very high-dimensional problems ($\approx 10^7$)
  by utilising the sparse nature of the precision matrix under Markov assumptions.
  Implementation wise, the sparsity is exploited by employing the sparsity possibilities made available
  by the \pkg{Eigen} \proglang{C++} linear-algebra library.
  The package and examples are available at \url{https://github.com/equinor/GraphSPME}.
}
\Plainkeywords{keywords, comma-separated, not capitalized, Java} %% without formatting
\Address{
  Berent Ånund Strømnes Lunde\\
  TDI EDT DSD\\
  Equinor\\
  Sandsliveien 90, 5254 Sandsli, Norway\\
  E-mail: \email{berl@equinor.com}\\
  Department of Mathematics\\
  Faculty of Mathematics and Natural Sciences\\
  University of Bergen\\
  E-mail: \email{berent.lunde@uib.no}\\
  \\~\\

  Feda Curic\\
  TDI EDT DSD\\
  Equinor\\
  Sandsliveien 90, 5254 Sandsli, Norway\\
  E-mail: \email{fcur@equinor.com}\\~\\
  Sondre Sortland\\
  TDI EDT DSD\\
  Equinor\\
  Sandsliveien 90, 5254 Sandsli, Norway\\
  E-mail: \email{sonso@equinor.com}
}
\begin{document}

%% include your article here, just as usual
%% Note that you should use the \pkg{}, \proglang{} and \code{} commands.

%\section[About Java]{About \proglang{Java}}
%% Note: If there is markup in \(sub)section, then it has to be escape as above.

\section{Introduction}\label{sec:introduction}

Efficient and accurate estimation of high-dimensional dependence structures 
is of increasing importance in computational statistics.
Machine-learning algorithms such as LDA and QDA require the computation of 
precision matrices of the feature vector \citep{hastie2009elements},
mixed-effect generalized linear models allow responses to share information through high-dimensional dependence modelling,
see e.g. the \pkg{mgcv} package \citep{wood2011mgcv},
similar dependence modelling is employed in packages such as \pkg{TMB} \citep{kristensen2015tmb} 
and \pkg{INLA} \citep{rue2009approximate} for
calculating the Laplace approximation when working with random effects.
Furthermore, ensemble-type filtering algorithms such as the ensemble Kalman filter \citep{evensen1994sequential, burgers1998analysis}, widely employed
in the fields of meteorology \citep{houtekamer2005ensemble}, oceanography \citep{evensen1994sequential}, and reservoir data assimilation \citep{aanonsen2009ensemble}
implicitly estimates the covariance, typically extremely high dimensional due to the spatio-temporal Gaussian random fields
involved in the problems.

The go-to estimator for dependence is the non-parametric sample covariance matrix.
This, as is well known, does not necessarily work well in high-dimensional problems
and in particular when the number of samples $n$ is smaller than the dimension $p$ where
the resulting estimator will be singular. 
Regularization of the estimation problem is thus necessary.
One common method is to use the Moore-Penrose generalized inverse, if a precision estimate is needed, retaining the non-zero 
eigenvalues of the sample covariance estimate. This is a common method in e.g. filtering methodology
and for LDA and QDA.
A different school of thought employs Stein-type shrinkage \citep{stein1956inadmissibility, james1961proc}:
a convex combination of the sample covariance matrix and some sparse (typically diagonal) target matrix \citep{ledoit2004well}.
Results for the amount of shrinkage, adaptive to the data, can be found in \citet{ledoit2004well,touloumis2015nonparametric},
thus computationally costly cross-validation \citep{stone1974cross} might be avoided 
depending on the target matrix.
This method is implemented in the \pkg{ShrinkCovMat} \proglang{R} package.
For the large class of ensemble filtering algorithms working on random fields, so-called localisation
is a necessary tool to overcome the problem of spurious correlations and corresponding noisy update-steps. 
Two common localisation methods, 
covariance localisation \citep{hamill2001distance, houtekamer2001sequential} 
and local analysis \citep{anderson2003local,evensen2003ensemble,ott2004local,hunt2007efficient}
both work in part on the correlation dependencies from the sample covariance matrix, using kernels to increasingly
dampen the strength of dependence as a function of distance between states.

The SPDE approach of \citet{lindgren2011explicit} linking the large class of continuous time 
models having SPDE specifications with the important class of indexed Gaussian Markov random fields (GMRF) \citep{rue2005gaussian}
motivates working directly with the precision matrix instead of the covariance.
This in part because of natural modelling assumptions using Markov properties, 
yielding a more parsimonious model and thus better statistical estimates, 
but also the computational savings in working with sparse matrices.
The sparsity is afforded due to the precision-parametrisation of the multivariate Gaussian with Markov properties.
Both the \pkg{TMB} and \pkg{INLA} packages exploit this sparsity
when solving problems involving high dimensional latent variable modelling.

Estimation of sparse precision matrices may be divided into the class of methods that jointly estimates parameters and the Markov properties,
and the class that estimates precision conditioned on knowing the zero and non-zero elements of the precision.
The former includes methods such as the well known graphical lasso \citep{friedman2008sparse}, and column-by-column 
methods \citep{yuan2010high,cai2011constrained,zhao2014calibrated,liu2017tiger}. 
For an overview, see \citet{fan2016overview}.
The latter class of methods typically involves the Gaussian likelihood, see e.g. \citet{hastie2009elements} or 
\citet{zhou2011high},
which involves iterative optimisation.
The method of \citet{le2022high} avoids this by utilising a column-by-column method inverting sample covariance matrices 
of dimension much smaller than the original $p$.
This scheme has both asymptotic results and is also computationally efficient. 
However, due to working on pure sample-covariance matrices, the numerical scheme may run into problems even for
moderate dimensions.
Furthermore, the resulting estimate is not symmetric, making e.g. the Cholesky decomposition inadmissible.

\pkg{GraphSPME} combines the precision estimation routine with respect to a graph in \citet{le2022high} with 
the automatic and adaptive shrinkage of sample covariance matrices in \citet{touloumis2015nonparametric}, and 
adds symmetry conversion to obtain computationally stable and fast estimates of precision matrices
given some graph or sparsity pattern.
The resulting estimates, that are guaranteed to be SPD, are possible to use with 
efficient factorisation routines such as the Cholesky decomposition when inverting or solving sparse 
linear systems in general.
It also adds methods for estimating the Markov order from data, given a graph of 1'st order neighbours.
The package is easy to use, provides fast computation, and works in very high dimensions.
Being easily available in \proglang{C++} as header only, and as \proglang{Python} and \proglang{R} packages
under PyPi and CRAN respectively, the package can be integrated in both machine-learning algorithms as well
as the large families of ensemble-filtering routines working on dynamical spatio-temporal models.

In Section \ref{sec:covariance} the ideas and concepts of asymptotic Stein-type shrinkage of 
the sample covariance estimate are introduced. Section \ref{sec:precision} covers in-depth 
the method of \citet{le2022high} that heavily influences this paper.
The covariance shrinkage method and the graphical precision method is combined in Section \ref{sec:software}
that also introduces methodology for estimating the Markov order of the data.
Usage of \pkg{GraphSPME} is illustrated through an example in Section \ref{sec:using},
while the properties are compared to competing methodologies in Section \ref{sec:case} working
on the auto-regressive process where all aspects such as sample size, dimension, and graphical structure
of the problem may be controlled and varied.
Section \ref{sec:discussion} concludes and discusses results.

\section{High-Dimensional covariance estimation with asymptotic Stein-type shrinkage}\label{sec:covariance}

It is well known that the sample covariance given by 
\begin{align}\label{eq:sample-covariance}
    \bm{S} = \frac{1}{n-1}\sum_{i=1}^n(\bm{x}_i - \overline{\bm{x}})(\bm{x}_i - \overline{\bm{x}})^\top
\end{align}
where $\bm{x}_i \in \mathbb{R}^{p\times1}$ are i.i.d random vectors and $\overline{\bm{x}}$ the sample mean, 
is not the best estimator when the number of samples $n$ is smaller than the number of parameterss $p$.
In addition to being singular for $n \leq p$, the sample covariance matrix can be poorly conditioned even when $n > p$, 
which means that inverting it amplifies estimation errors \citep{ledoit2004well}.
Estimators of covariance are important in many applications and much work has been done to improve them.
One promising idea, proposed by \citet{ledoit2004well}, is to devise estimators based on shrinkage, where the following convex combination of the sample covariance and some target matrix is used as estimator:
\begin{align}\label{eq:stein-covariance-shrinkage}
    \bm{S^\star} = (1 - \lambda)\bm{S} + \lambda \bm{T}.
\end{align}
Here, $\bm{T}$ is a target matrix, for the time being equalling 
$\nu \bm{I}_p$ 
where $\bm{I}_p$ is the $p \times p$ identity matrix and $\lambda$ and $\nu$ are chosen to minimise the risk function $E[|| \bm{S^\star} - \bm{\Sigma} ||_F^2 ]$ -- the expected Frobenius norm on the difference between the population and estimated covariance.
We now give a short intuitive introduction to Stein-type shrinkage and how it leads to results used in \pkg{GraphSPME}.

\citet{stein1956inadmissibility} showed that the sample mean is inadmissible (there exist better estimators) when $p \geq 3$.
\citet{james1961proc} proposed a new and better estimator, called the James-Stein estimator, that in essence shrinks individual sample means toward some target.
The target is typically, but not necessarily, chosen to be the grand mean or average of averages.
How much each individual sample mean is shrunk is given by a shrinkage factor.
To formalise, let $\bm{X} \in \mathbb{R}^{n \times p}$ be a matrix of random variables, with sample mean of the $j$'th variable equal $\hat{\mu}_j = \frac{1}{n}\sum_{i=1}^n x_{ij}$, and average of averages equal $\tilde{\mu} = \frac{1}{p} \sum_{j=1}^p \hat{\mu}_j$.
The James-Stein estimator can then be written as
\begin{align}
    \mu_j^\star = (1 - \lambda)\tilde{\mu} + \lambda \hat{\mu}_j
\end{align}
which is analogous to Equation \ref{eq:stein-covariance-shrinkage}.
The covariance estimator proposed by \citet{ledoit2004well} can in similar vein be interpreted as shrinkage towards a grand mean, but now towards $\frac{tr(\bm{S})}{p}$ (average sample variance) instead of the sample mean.
Another interesting interpretation as given in \citet{ledoit2004well}, is that of shrinking the eigenvalues of the sample covariance matrix towards their grand mean.

One issue of the covariance shrinkage estimator \eqref{eq:stein-covariance-shrinkage}, is that the optimal shrinkage 
\begin{align}\label{eq:optimal-shrinkage-theoretical}
    \lambda = \frac{E\left[|| \bm{S} - \bm{\Sigma}||_F^2\right]}{E[|| \bm{S} - \nu \bm{I}_p||_F^2]}
\end{align}
proposed by \citet{ledoit2004well} is not a bona fide estimator, as it depends on the true and unobservable covariance matrix.
They solve this by using general asymptotics (where $p$ goes to infinity at the same speed as $n$) to construct consistent estimators.
\citet{touloumis2015nonparametric} builds upon this work 
and is able to derive estimators for $\lambda$ under a general non-parametric framework.
Firstly, it is obtained that 
\begin{align}\label{eq:optimal-shrinkage-theoretical-asymptotical}
    \lambda^\star = \frac{tr(\bm{\Sigma}^2) + tr^2(\bm{\Sigma})}{Ntr(\bm{\Sigma}^2) + \frac{p-N+1}{p} tr^2(\bm{\Sigma})},
\end{align}
by first expanding the expectations in the numerator and denominator of \eqref{eq:optimal-shrinkage-theoretical} 
and retaining only quantities that are asymptotically non-negligible.
To create a consistent estimator of \eqref{eq:optimal-shrinkage-theoretical-asymptotical}, 
parametervectorthe quantities are replaced by consistent estimators constructed using U-statistics.
The proposed estimator as given in \citet{touloumis2015nonparametric} is 
\begin{align}\label{eq:optimal-shrinkage-asymptotic}
    \hat{\bm{S}}^\star = (1 - \hat{\lambda})\bm{S} + \hat{\lambda}\hat{\nu}\bm{I}_p 
\end{align}
with
\begin{align}
    \hat{\lambda} = \frac{Y_{2N} + Y_{1N}^2}{NY_{2N} + \frac{p-N+1}{p}Y_{1N}^2}
\end{align}
and $\hat{\nu} = Y_{1N}/p$.
$Y_{1N}$ and $Y_{2N}$ are constructed using standard results from the theory of U-statistics as $Y_{1N} = U_{1N} - U_{4N}$ and $Y_{2N} = U_{2N} - 2U_{5N} + U_{6N}$.

Similar results follow for alternative target matrices $\bm{T}$ and are constructed in the same manner.
\pkg{GraphSPME} implements the covariance estimator of \citet{touloumis2015nonparametric} for a target matrix
containing the sample variances on the diagonal.
This is available for users, but also employed under-the-hood in precision matrix estimation
further elaborated on in Section \ref{sec:software}.

\section{High-dimensional precision estimation with known Markov properties}\label{sec:precision}
The covariance parametrisation of dependence and corresponding estimates of Section \ref{sec:covariance} are efficient
when no more structure on the data generating process is known.
However, when e.g. dimensions are known to be conditionally independent, other parametrisations and corresponding estimates may be more efficient.
In particular, parametrisations filtering the family of distributions to distributions that factor according to the conditional independence 
\citep{bishop2006pattern} are particularly useful.
We will here discuss estimation of the precision matrix under Gaussian Markov Random Field assumptions.

Let $\mathcal{G}=(\mathcal{V},\mathcal{E})$ be a graph with vertices $\mathcal{V}$ and edges $\mathcal{E}$. 
A random vector
$\bm{x}\in R^d$ 
is a Gaussian Markov Random Field (GMRF) \citep{rue2005gaussian} with respect to the graph
$\mathcal{G}=(\{1,\ldots,d\},\mathcal{E})$, 
with mean $\mu$ and symmetric positive definite (SPD) precision matrix $\Lambda$ if
\begin{align}
    p(x)=(2\pi)^{-\frac{d}{2}}\sqrt{|\Lambda|}\exp\left(-\frac{1}{2}(\bm{x}-\bm{\mu})^\top\Lambda (\bm{x}-\bm{\mu})\right)
\end{align}
and
\begin{align}
    \Lambda_{i,j} \neq 0 \Leftrightarrow (i,j) \in \mathcal{E} \forall i\neq j.
\end{align}
One of the advantages of the mean-precision parametrisation versus the mean-covariance parametrisation 
of the GMRF is the intrinsic filtering of the family of Gaussian distributions.
Only the Gaussian distributions satisfying factorisation of the joint distribution 
due to conditional independence are considered under a known sparsity pattern of the precision.
Necessarily, estimation of the precision with known zeroes will imply searching only over 
distributions that satisfy specified Markov properties corresponding to some pre-specified graph $\mathcal{G}$.
This is, of course, highly beneficial to e.g. spatio-temporal modelling that frequently
utilises said properties.

%> Figure, graph, GMRF, factorization, precision sparsity-pattern

The graphical lasso algorithm \citep{friedman2008sparse} famously penalises dense precision
and searches for an optimal (w.r.t. the Gaussian likelihood)
sparsity pattern in the space of SPD matrices. 
Similar methods estimate the precision column-by-column 
\citep{yuan2010high,cai2011constrained,zhao2014calibrated,liu2017tiger}
parametervectorby exploiting the relationship between the conditional distributions from the multivariate normal
and linear regression to employ lasso-type regression algorithms that enforce sparsity.
Common for all these methods is that they search for the sparsity pattern and the corresponding 
graph, without it being specified pre-estimation.

The problem of estimating the precision matrix under a known sparsity pattern or graphical structure
has received less attention than the problem of jointly estimating non-zero and zero elements of 
the precision as discussed above, 
and most methods require the Gaussian likelihood \citep{hastie2009elements,zhou2011high} in tandem
with iterative optimisation. 
The method of \citet{le2022high} is a computationally efficient column-by-column method that explicitly
estimates the non-zero elements $w_{j1}$ of column $w_j = \Lambda_{.j}$ of the precision using 
block-sample covariance matrices.
Let $B_j$ be a matrix of zeroes and ones so that $\bm{B}_jw_{j1}=w_j$.
Accordingly, let $\bm{x}_{i\bm{B}_j}=\bm{B_j}^\top\bm{x}_i$ denote the sub-vector of $\bm{x}_i$ where relevant dimensions are "picked-out"
by $\bm{B}_j$.
Correspondingly, 
$S_{\bm{x},j}=\frac{1}{n-1}\sum_i (\bm{x}_{i\bm{B}_j}-\overline{\bm{x}_{\bm{B}_j}})(\bm{x}_{i\bm{B}_j}-\overline{\bm{x}_{\bm{B}_j}})^\top$ denotes the $j$-th block-sample covariance matrix and where $\overline{\bm{x}_{\bm{B}_j}}$ denotes the sample mean of $\{\bm{x}_{i\bm{B}_j}\}_{i=1}^n$.
Then \citet{le2022high} estimates the $j$-th column-vector of the precision by
\begin{align}
    \hat{\bm{w}}_j = 
    \bm{B}_j
    S_{\bm{x},j}^{-1}
    \bm{B}_j^\top \bm{e}_j
\end{align}
where $\bm{e}_j$ is the $j$-th column of the $p\times p$ identity matrix $\bm{I}_p$.

Under relatively mild assumptions of positive bounded eigenvalues of the population covariance 
and that the $||\bm{\Sigma}||_1$ norm is bounded, \citet{le2022high} establishes asymptotic
normality on the estimated non-zero elements of column $j$, $\hat{\bm{w}}_{j1}$.
The numerical scheme is not guaranteed to succeed, as even a block-sample covariance matrix
can be singular, depending on the connectivity of $\mathcal{G}$ and the number of observations $n$.
Furthermore, the estimated precision is generally not symmetric due to the 
column-by-column nature of the routine.
The preceding section seeks to build on the method of \citet{le2022high} by targeting 
these issues, and propose an efficient implementation in the \pkg{GraphSPME} library.

%\begin{enumerate}
%    \item conditional independence, GMRF assumptions leads to precision matrix parametrisation
%    \item GMRF definition
%    \item Literature review of precision estimates: estimate graph (graphical lasso, col-by-col graphical danzig, TIGER), estimate wrt graph (Gaussian-ML-estimation, Le2021)
%    \item The method of Le2021 in detail
%    \item Asymptotic propertiess and not SPD
%\end{enumerate}

\section{Software implementation and innovations} \label{sec:software}

\begin{figure}
    \centering
    \begin{tikzpicture}
    %\usetikzlibrary{shapes.misc, positioning}
    \tikzstyle{vertex}=[circle,draw=black,fill=black!10,minimum
    size=16pt,inner sep=1pt]
    \tikzstyle{language}=[circle,draw=black,fill=gray!10,minimum
    size=26pt,inner sep=2pt]
    \tikzstyle{library}=[rounded rectangle,draw=black, minimum size=20pt, inner sep=2pt]
    % vertices
    \node[library] (eigen) at (0,0) {\textbf{Eigen}};
    \node[language] (cpp) [below=1cm of eigen] {\textbf{C++}};
    \node [right=0.1cm of cpp] {header-only};
    \node[library] (pybind11) [above right=0.5cm and 1cm of cpp] {\textbf{pybind11}};
    \node[library] (rcpp) [below right=0.5cm and 1cm of cpp] {\textbf{Rcpp}};
    \node[language] (python) [right=0.6cm of pybind11] {\textbf{Python}};
    \node [above of=python]{PyPi};
    \node[language] (R) [right=0.8cm of rcpp] {\textbf{R}};
    \node [below of=R]{CRAN};
    \node[library] (scipy) [right=1.0cm of python] {\textbf{scipy.sparse}};
    \node[library] (Matrix) [right=1.0cm of R] {\textbf{Matrix}};
    % edges
    \path[draw,thick,->] (eigen) -- (cpp);
    \path[draw,thick,->] (cpp) -- (pybind11);
    \path[draw,thick,->] (cpp) -- (rcpp);
    \path[draw,thick,->] (pybind11) -- (python);
    \path[draw,thick,->] (rcpp) -- (R);
    \path[draw,thick,dashed,->] (scipy) -- (python);
    \path[draw,thick,dashed,->] (Matrix) -- (R);
    \end{tikzpicture}
    \caption{
    Overview of the \pkg{GraphSPME} package. The package is implemented as 
    header-only in \proglang{C++} \citep{stroustrup2000c++}, 
    and exposed to \proglang{R} \citep{Rlanguage2018} (available at CRAN) 
    and \proglang{Python} \citep{PythongLanguage} (available at PyPi) via
    the packages \pkg{Rcpp} \citep{rcpp2011R} and \pkg{pybind11} \citep{pybind11} respectively.
    Utilising \pkg{GraphSPME} requires sparse matrices, which implementation may
    be found in e.g. \pkg{scipy.sparse} \citep{2020SciPy-NMeth} for \proglang{Python} 
    or \pkg{Matrix} \citep{MatrixRPackage} for \proglang{R}.
    }
    \label{fig:graphspme-package}
\end{figure}
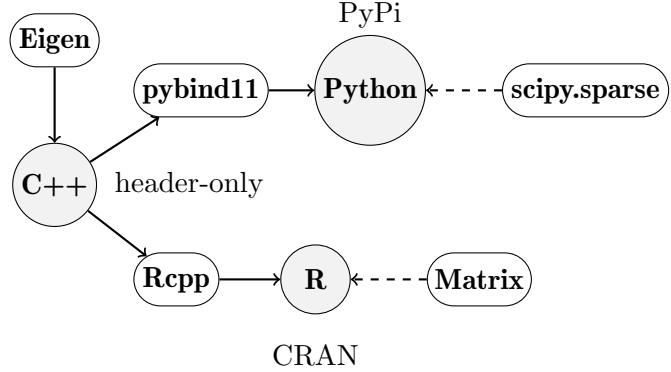

At its core, \pkg{GraphSPME} implements sparse precision matrix estimation with respect 
to a known graphical structure as discussed in Section \ref{sec:precision}.
In addition to the method of \citet{le2022high}, the automatic asymptotic covariance shrinkage
estimate of \citet{touloumis2015nonparametric} and introduced in Section \ref{sec:covariance} is also 
implemented and exposed to users of the package.
This allow modifying the sparse precision matrix estimate, 
to resolve the issue of potential singular block-sample covariance matrices
that are repeatedly inverted. 
It is reasonable to also expect slightly more efficient estimates of the full 
precision matrix in comparison to the original method that is without any shrinkage.
Finally, a symmetry conversion
\begin{align}
    \hat{\bm{\Lambda}}=\frac{1}{2}\left(\tilde{\bm{\Lambda}}+\tilde{\bm{\Lambda}}^\top \right)
\end{align}
is imposed on the final estimate.
This does however not degrade the result: \citet{cai2011constrained} shows that 
$\hat{\bm{\Lambda}}$
achieve the same rate of convergence as
$\tilde{\bm{\Lambda}}$
asymptotically, and asymptotic results as presented in \citet{le2022high} may therefore
still be admissible.

\subsection{Markov-order estimation}
\label{subsec:markov-order estimation}

In real applications, the neighbourhood structure is oftentimes known, while the 
Markov order might be unknown. 
It may be exactly imposed through modelling choices, such as assuming efficient markets
yielding a first order condition on prices, or the modelling assumptions are flexible 
enough to allow for a structurally-given but specifically unknown dependence structure.
%Such cases may occur when modelling stochastic processes using SPDEs \citep{lindgren2011explicit} -- the 
%first-order Euler scheme will typically be 1'st order Markov, but due to integration
%of the SPDE, a higher order Markov order can be induced into the system between 
%observed time-points. <double check this argument>

The argument above is the reason for the separation of graphical neighbourhood 
structure and the Markov order in \pkg{GraphSPME}. 
In the function estimating sparse precision matrices, both the neighbourhood graph
(identical to a 1'st order Markov order graph) 
and the integer Markov order (potentially estimated as above) can be provided
as arguments.

For the cases of an unknown Markov order but a given neighbourhood structure,
\pkg{GraphSPME} implements a function for estimating the exact nature of the order.
The underlying method starts at Markov order 0 (independence), iteratively increments 
the order while running tests, and stops when some ideal order is found.
During iterations, the average Gaussian likelihood (and generally the quasi-likelihood) 
\begin{align}
    l(\bm{\Lambda}) 
    = \frac{1}{2}\left( \texttt{tr}\left(\bm{S}\bm{\Lambda}\right)-\log\left|\bm{\Lambda}\right|\right)
\end{align}
is calculated.
This function is monotonically decreasing in Markov order, due to the hierarchical 
nature of the different models. 
To decide on some order, penalization must be added.
\pkg{GraphSPME} utilizes the AIC value, the number of estimated parameters, scaled by
$n^{-1}$ as we are working with an average likelihood. \citep{akaike1974new}
Let $m$ be the number of columns in the data $\bm{X}$, and $l$ be the number of non-zero entries 
in the estimated precision matrix $\hat{\bm{\Lambda}}$.
The penalized quasi average likelihood is then given as
\begin{align}
    l_{\text{aic}}(\hat{\bm{\Lambda}}) 
    = \frac{1}{2}\left( \texttt{tr}\left(\bm{S}\hat{\bm{\Lambda}}\right)-\log\left|\bm{\hat{\Lambda}}\right| \right)
    + \frac{1}{2n}(l+m).
\end{align}
Since $\frac{1}{2}(l+m)$ gives the number of parameters due to the symmetry 
of the precision.
The arguments for using the AIC is its computational speed and stability over other 
information criteria \citep{takeuchi1976distribution,murata1994network},
or computationally costly cross-validation \citep{stone1974cross}. 
In addition, the precision estimating routine 
employing the block-sample covariance matrices being only a scaling factor 
away from the maximum likelihood estimate, making the AIC asymptotically amenable.

Note that since datasets typically are finite, it can occur that estimating
only a sub-vector of the conceptually true parameter-vector can yield a better result (in expectation on test data)
than estimating the full parameter-vector.
This is due to the extra variance that is induced by increasing the dimension on 
the space of parameters.
The workflow utilising estimation of Markov order is illustrated in Section \ref{sec:using}.

\subsection{Implementation details}
\label{subsec:implementation details}

Figure \ref{fig:graphspme-package} illustrates the overall structure and dependencies of 
\pkg{GraphSPME}.
The core functionality in the package is implemented in \proglang{C++}, relying
heavily on the time-tested fast linear-algebra library \pkg{Eigen}, 
and in particular the matrix classes afforded by its sparsity module.
It is implemented as header-only, and therefore easily included by other packages.
Bindings are also created to both \proglang{Python} and \proglang{R}, for easy
availability to users of these languages.
For \proglang{Python} this is done via \pkg{pybind11},
and with \pkg{Rcpp} to \proglang{R}.
Working with \pkg{GraphSPME} in these langauges requires libraries making
sparse matrices available to represent graphs and return values.
For \proglang{Python}, \pkg{GraphSPME} is available through PyPi, and through
CRAN for \proglang{R} users.

\section{Using the GraphSPME package} \label{sec:using}
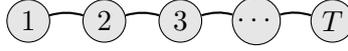
\begin{figure}
    \centering
    \begin{tikzpicture}
    \tikzstyle{vertex}=[circle,draw=black,fill=black!10,minimum
    size=16pt,inner sep=1pt]
    \node[vertex](1) at (0,0) {1};
    \node[vertex](2) [right of=1] {2};
    \node[vertex](3) [right of=2] {3};
    \node[vertex](D) [right of=3] {$\cdots$};
    \node[vertex](T) [right of=D] {$T$};
    \path[draw,thick,-] (1) to[out=15,in=165] (2);
    \path[draw,thick,-] (2) to[out=15,in=165] (3);
    \path[draw,thick,-] (3) to[out=15,in=165] (D);
    \path[draw,thick,-] (D) to[out=15,in=165] (T);
    \end{tikzpicture}
    \caption{
    The neighbourhood structure of the mixed-effect AR-$p$ model, corresponding to 
    the graph created for a 1'st order Markov process.
    }
    \label{fig:mixed-effect-ar1}
\end{figure}
\begin{figure}[t!]
    \centering
    \includegraphics[width=0.7\textwidth]{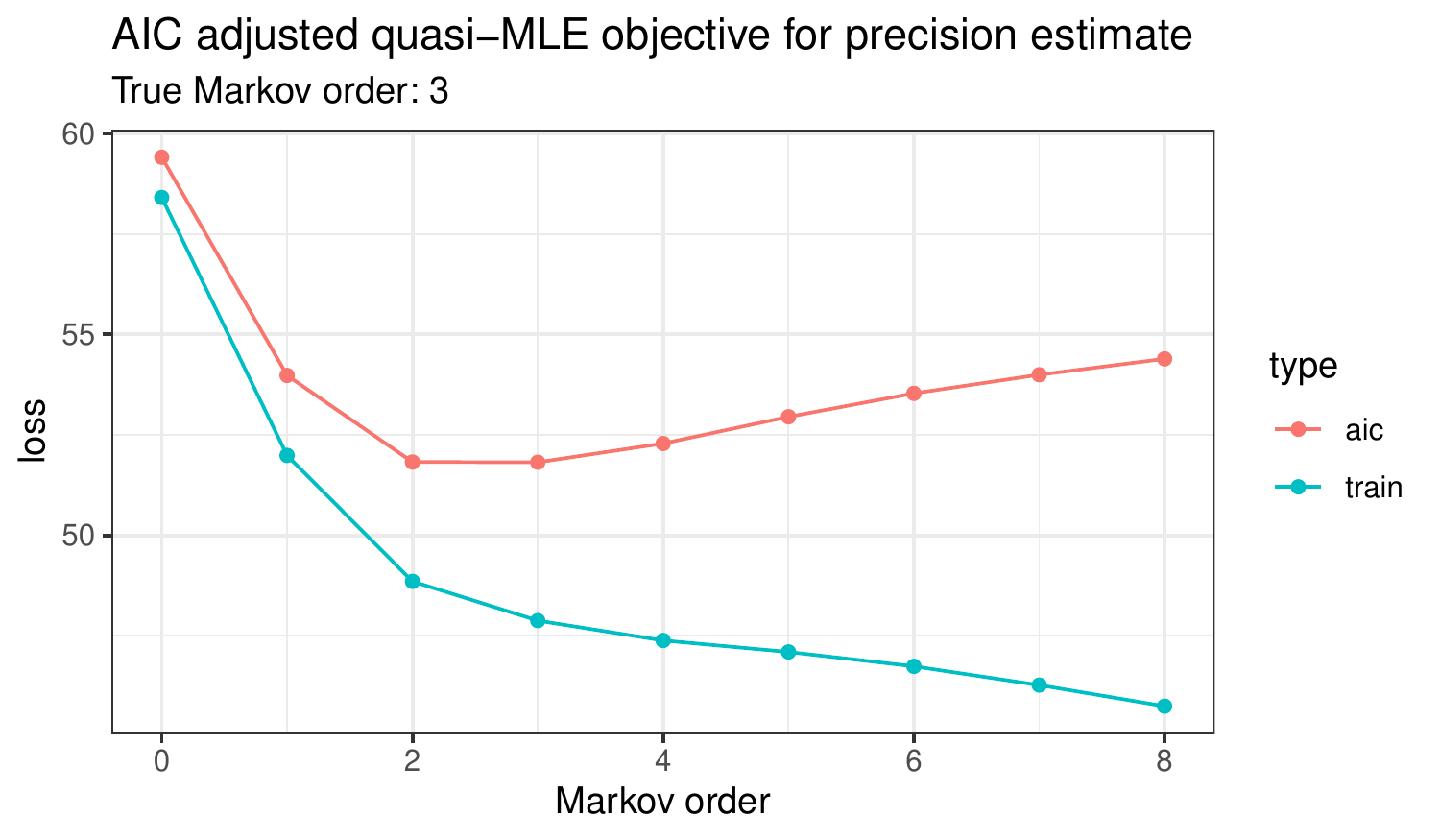}
    \caption{
    Results from estimating the Markov order for the mixed-effect AR-$p$ model \eqref{eq:mixed-effect-arp}.
    The "train" results are from the average Gaussian loss implemented in \code{prec\_nll}, while the 
    "aic" penalization is obtained through the \code{prec\_aic} implementation.
    Both taking a precision estimate obtained through the \code{prec\_sparse} function using 
    various degrees of Markov order ($x$-axis).
    }
    \label{fig:markov-order-estimation}
\end{figure}

The goal of \pkg{GraphSPME} is to provide dependence estimates with respect
to a graph with some degree of connectivity, potentially either fully connected or disconnected.
For a fully connected graph, it makes sense to directly estimate the covariance matrix
and not the precision matrix. 
For this edge case (but also very typical), \pkg{GraphSPME} provides the \code{cov\_shrink\_spd}
which returns the shrinkage covariance estimate of \citet{touloumis2015nonparametric} for a sample-variance diagonal
target matrix presented in Section \ref{sec:covariance}.
For all other cases of a not fully connected graph, it is sensible to estimate the precision matrix
due to the sparsity afforded by missing vertices between edges relative to the fully connected variant.
In such situations, the function \code{prec\_sparse} is implemented, that returns the 
precision estimate of \citet{le2022high} potentially with the shrinkage and symmetry adjustments described in 
Section \ref{sec:software}.
\code{prec\_sparse} requires the $n\times p$ dataset as input, along with a graph describing the
neighbourhood structure, and an integer Markov order (defaults to 1).
If the Markov order is unkown, \pkg{GraphSPME} finally implements the AIC for the precision estimate
in \code{prec\_aic} for a given training dataset and corresponding precision estimate with some 
assumed Markov order.
This may then be employed to estimate the degree of connectivity in the graph, given a neighbourhood 
structure.

Following is a walk-through of the above mentioned functions of \pkg{GraphSPME}.
The full code may be found at 
\url{https://github.com/equinor/GraphSPME/tree/main/GraphSPME-examples/mixed_effect_arp.R}.
Consider the $i$'th realisation of  $i=1:n$-realisations of a random-effect AR-$p$ model
\begin{align}\label{eq:mixed-effect-arp}
    x_{i,t} = u_t\sum_{j=1}^p \psi_j x_{i,t-j} + \epsilon_{i,t},~~
    \sum_j \psi_j = 1,~~
    u_t\sim U(0,1),~~
    \epsilon_{i,t}\sim N(0,1)
\end{align}
where the random effects $u_t$ are indexed by time $t$.
Assume that we have a dataset of $n=100$ independent realisations of this model with final time $T=100$.
For \pkg{GraphSPME} it is necessary to work with a defined neighbourhood structure.
In this model, the neighbourhood structure correspond to a 1'st order Markov structure
as illustrated in Figure \ref{fig:mixed-effect-ar1}.
To pass this along to \code{prec\_sparse} we create the corresponding sparse 
matrix $G$ with ones if $i$ is a neighbour of $j$.
\begin{CodeChunk}
\begin{CodeInput}
R> G <- bandSparse(100, 100, 
                (-1):1, 
                sapply((-1):1, function(j) rep(1,100-abs(j)))
   )
\end{CodeInput}
\begin{CodeOutput}
100 x 100 sparse Matrix of class "dgCMatrix"
[1,] 1 1 . . . . . 
[2,] 1 1 1 . . . . 
[3,] . 1 1 1 . . . 
[4,] . . 1 1 1 . . 
[5,] . . . 1 1 . . 
..................
\end{CodeOutput}
\end{CodeChunk}
Note that the function \code{bandSparse} from the \pkg{Matrix} package is used to 
be allowed to work with sparse matrices on the \proglang{R} side, see also 
Figure \ref{fig:graphspme-package}.
In the case when $p$ is unknown, i.e. the Markov order, the user can employ the 
\code{prec\_aic} function for different precision estimates where the 
Markov order is incremented through the \code{markov\_order} argument.
\begin{CodeChunk}
\begin{CodeInput}
R> nll <- aic <- numeric(16)
R> for(p in 0:15){
       Prec_est <- prec_sparse(xtr, G, markov_order=p)
       nll[p+1] <- prec_nll(xtr, Prec_est)
       aic[p+1] <- prec_aic(xtr, Prec_est)
   }
\end{CodeInput}
\end{CodeChunk}
The results are visualised in Figure \ref{fig:markov-order-estimation}. 
The routine correctly identifies the AR-3 model, even with relatively little data,
making the results very close to the more parsimonious AR-2 model.
Given that the Markov-order is identified, the user may produce the final precision estimate
through again employing the \code{prec\_sparse} function.
\begin{CodeChunk}
\begin{CodeInput}
R> Prec <- prec_sparse(xtr, G, 3)
\end{CodeInput}
\end{CodeChunk}
\begin{CodeOutput}
100 x 100 sparse Matrix of class "dgCMatrix"
[1,]  1.2807 -0.4191 -0.1180  0.0653  .      .       .     
[2,] -0.4191  0.9841  0.0659 -0.2544 -0.112  .       .     
[3,] -0.1180  0.0659  1.1742 -0.3586 -0.276  .       .     
[4,]  0.0653 -0.2544 -0.3586  1.1861 -0.319  .       .     
[5,]  .      -0.1123 -0.2765 -0.3190  0.819  .       .     
......................................................
\end{CodeOutput}
Because the precision estimate is non-parametric, estimates of 
the fixed effects $\bm{\psi}$ and the random effects $\bm{u}$ may then be found
through estimating the conditional expectation of all data-points using
the precision estimate
\begin{align}
    E[x_i|x_{ne(i)}] = -\bm{\Lambda}_{i,i}^{-1} \sum_{j\in ne(i)}\bm{\Lambda}_{i,j}x_j,
\end{align}
which assumes the unconditional mean to be zero, 
and then relating such estimates to the predictive formula \eqref{eq:mixed-effect-arp}
involving both fixed and random effects.

%\begin{enumerate}
%    \item Geographical visual data
%    \item Natural neighbourhood function
%    \item Create neighbourhood graph 
%    \item Estimate Markov order
%    \item Estimate Precision matrix
%    \item Plot precision?
%\end{enumerate}

\section{High-Dimensional big-data case study}\label{sec:case}

\begin{figure}[t!]
    \centering
    \includegraphics[width=0.9\textwidth]{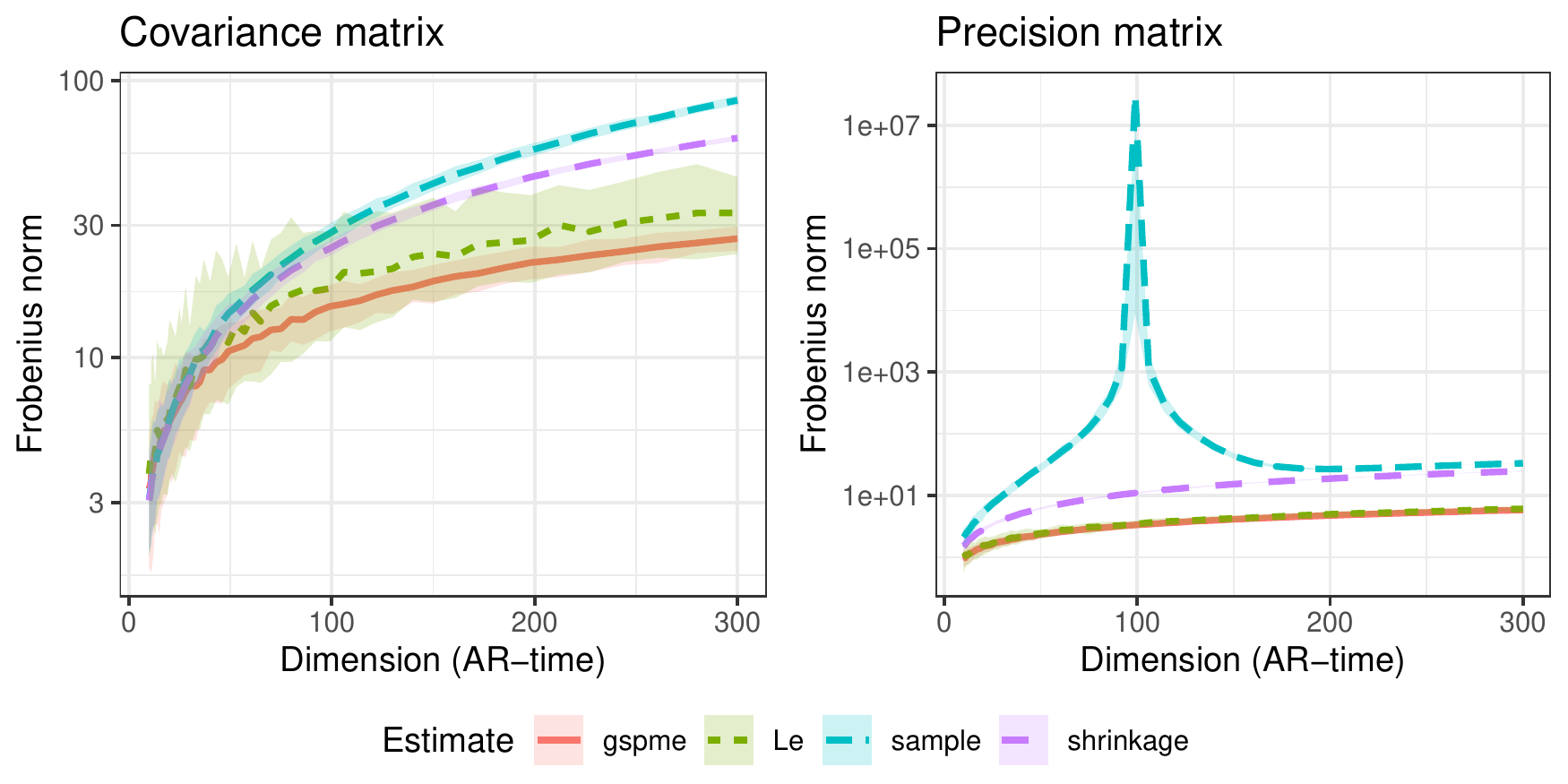}
    \caption{
    The Frobenius norm ($y$-axis, log-scale) evaluated on the difference between population 
    covariance (left) or precision (right) matrices and corresponding
    statistical estimates, explained in the start of Section \ref{sec:case}.
    The statistical process is an AR-1 process with sample size fixed at $n=100$, and
    parameter $\psi=0.8$, while the time or dimension of the problem ranges from
    10 to 300 ($x$-axis).
    The solid line is the average over 100 Monte Carlo simulations, while the 
    transparent confidence bands yields the empirical 0.05 and 0.95 empirical quantiles.
    }
    \label{fig:fbnorm-vs-dim}
\end{figure}

\begin{figure}[t!]
    \centering
    \includegraphics[width=0.9\textwidth]{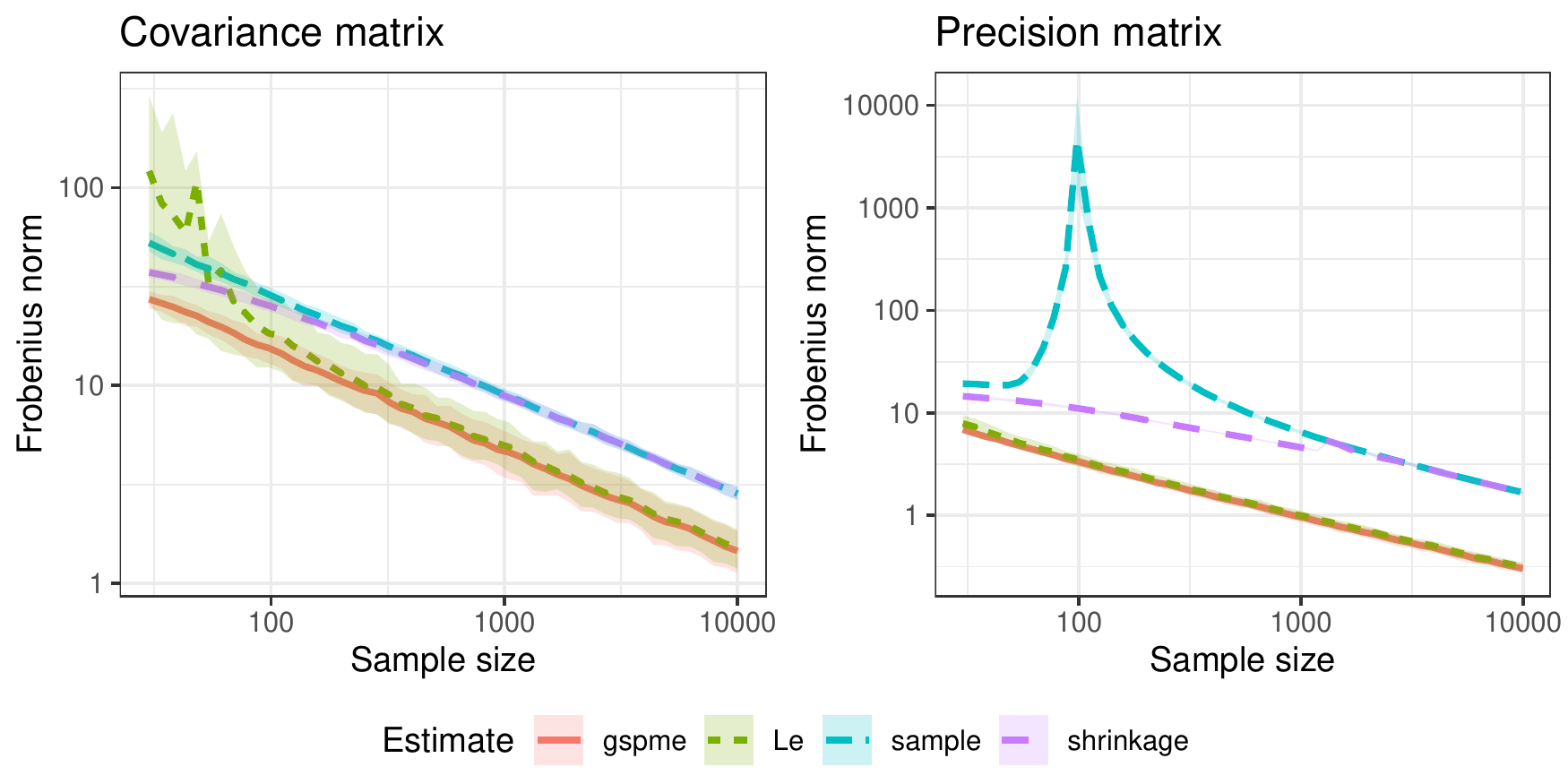}
    \caption{
    The Frobenius norm ($y$-axis, log-scale) evaluated on the difference between population 
    covariance (left) or precision (right) matrices and corresponding
    statistical estimates, explained in the start of Section \ref{sec:case}.
    The statistical process is an AR-1 process with dimension or time fixed at $T=100$, and
    parameter $\psi=0.8$, while the sample size ranges from
    30 to 10000 ($x$-axis, log-scale).
    The solid line is the average over 100 Monte Carlo simulations, while the 
    transparent confidence bands yields the empirical 0.05 and 0.95 empirical quantiles.
    }
    \label{fig:fbnorm-vs-sample-size}
\end{figure}

\begin{figure}[t!]
    \centering
    \includegraphics[width=0.9\textwidth]{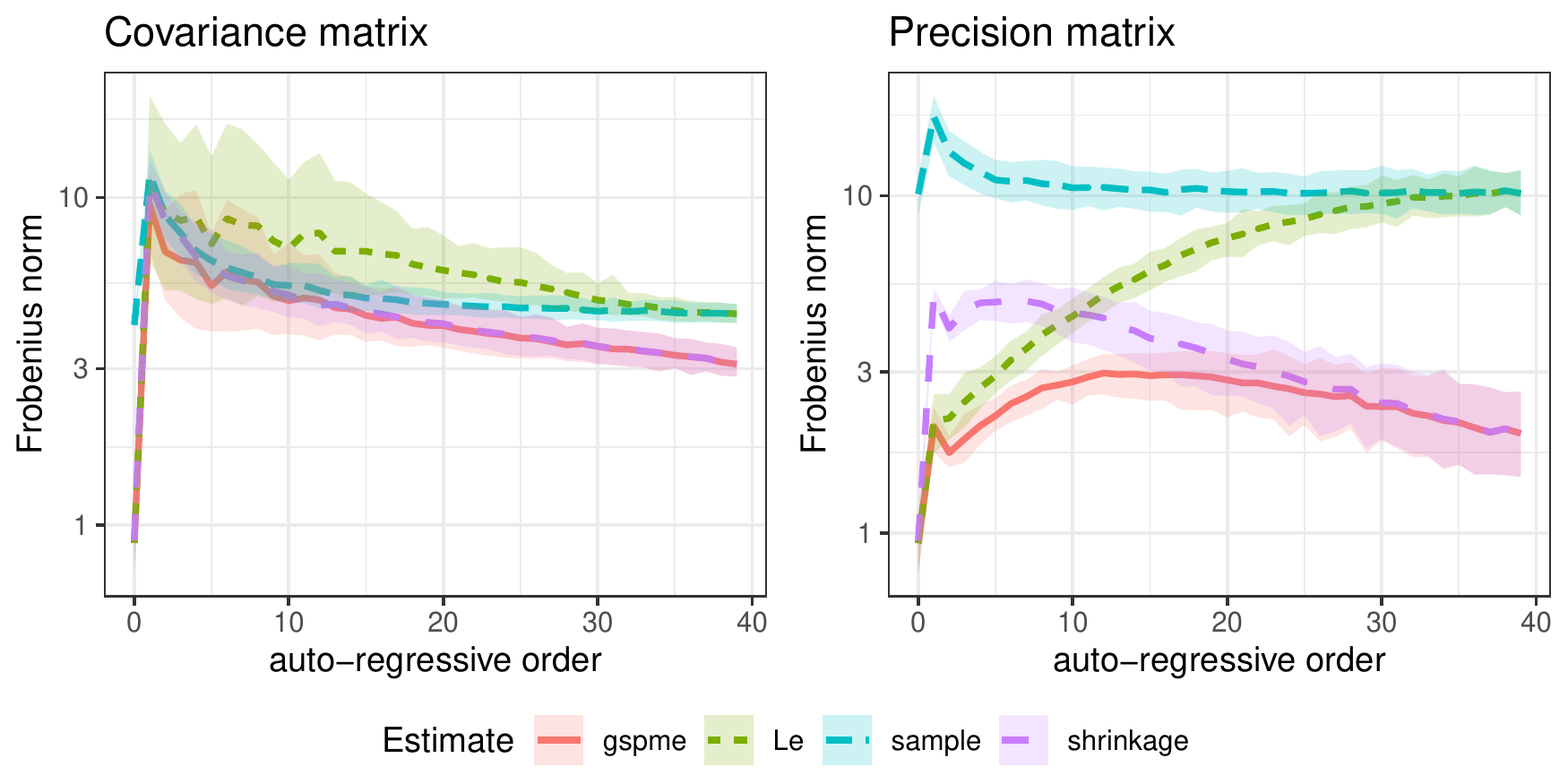}
    \caption{
    The Frobenius norm ($y$-axis, log-scale) evaluated on the difference between population 
    covariance (left) or precision (right) matrices and corresponding
    statistical estimates, explained in the start of Section \ref{sec:case}.
    The statistical process is an AR-$p$ process with sample size fixed at $n=100$,
    time fixed to $T=40$, while the auto-regressive order ($p$) of the process ranges
    from 0 (giving independent white noise with a fully disconnected graph) 
    to a fully-connected graph at $p=T-1=39$ ($x$-axis).
    All $\psi$'s are equally set to $0.8/p$.
    The solid line is the average over 100 Monte Carlo simulations, while the 
    transparent confidence bands yields the empirical 0.05 and 0.95 empirical quantiles.
    }
    \label{fig:fbnorm-vs-ar-order}
\end{figure}

The dependency estimates implemented in \pkg{GraphSPME} are tested
against one another and also versus the sample covariance matrix as a default benchmark
that all estimates should improve upon.
To control for uncertainty, we perform 100 Monte Carlo simulations for each experiment.
The results are presented visually in figures \ref{fig:fbnorm-vs-dim}, \ref{fig:fbnorm-vs-sample-size}, and \ref{fig:fbnorm-vs-ar-order}.
Here, the estimate resulting from the estimation routine implemented 
with \pkg{GraphSPME} explained in Section \ref{sec:software} is denoted "gspme", 
the estimate of \citet{le2022high} is denoted "Le", 
the shrinkage estimate of \citet{touloumis2015nonparametric} is denoted "shrinkage",
and the benchmark sample covariance estimate \eqref{eq:sample-covariance} is denoted "sample".

We employ the AR-$p$ process having known covariance and precision matrices, so that 
we may use e.g. the Frobenius norm to measure the accuracy of the estimates.
We vary both the sample size, $n$, the dimension $T$, and the auto-regressive order $p$,
to test all aspects of the dependence estimates given in \pkg{GraphSPME}. 
As an example, let 
\begin{align}
    x_t=\phi x_{t-1}+\epsilon_t,~ x_1\sim\mathcal{N}\left(0,\frac{1}{1-\phi^2}\right),~\epsilon_t\sim \mathcal{N}(0,1)
\end{align}
be an AR-1 process.
Then, the joint distribution of $\bm{x}=[x_1,\cdots,x_T]^\top$ has dense covariance given by
\begin{align}
\bm{\Sigma} &=
\begin{bmatrix}
B(1,1) & \cdots & B(1,T)\\
\vdots & \ddots & \vdots \\
B(T,1) & \cdots & B(T,T) 
\end{bmatrix},~
\text{where }
B(i,j)=\frac{\phi^{|i-j|}}{1-\phi^2},
\end{align}
while the equivalent precision matrix is sparse, and given by
\begin{align}
\Lambda =
\begin{bmatrix}
1 & -\phi \\
-\phi & 1+\phi^2 & -\phi \\
& \ddots & \ddots & \ddots \\
& & -\phi & 1+\phi^2 & -\phi \\
& & & -\phi & 1
\end{bmatrix}.
\end{align}
Similar results hold for the AR-$p$ process - the covariance is dense with elements given by a 
known autocovariance function, while the precision matrix will be sparse having a band-sparse structure.
It is therefore ideally suited to study the properties of dependence estimates with respect to some graph.
We use the \code{ARMA.var} function of the \pkg{ts.extend} \proglang{R}-package to calculate the 
exact population covariance matrix, and then invert it to obtain the population precision matrix.
The full code can be found at
\url{https://github.com/equinor/GraphSPME/tree/main/GraphSPME-examples/arp_dimension_study.R}.

% dimension increase
Figure \ref{fig:fbnorm-vs-dim} contains the results from fixing sample size $n=100$ of the AR-1 ($\psi=0.8$)
while varying the dimension of the joint distribution by increasing time $T=10$ to $T=300$ and 
then estimating dependence.
Note the log $y$-axis.
Due to the fixed sample size, we expect the difference to the population quantities to be increasing 
in dimension, as the datapoints take up less and less space in a high-dimensional space 
(manifesting the curse of dimensionality, as there is decreasing information per unit volume).
This is confirmed for all estimation procedures.
Both the covariance and the precision figures give the same picture of performance: 
Looking at the solid lines representing the averages,
the \pkg{GraphSPME} algorithm dominates the other three, with Le being almost indistinguishable 
for the precision estimates, but can be seen to vary (having a much broader confidence band) 
and be much more unstable when inverted to a covariance estimate. 
This is likely due to the pure block-sample covariance matrices of the method, that are automatically regularised with adaptive shrinkage in \pkg{GraphSPME}.
The advantage of \pkg{GraphSPME} and Le routines over the two covariance estimation routines should come as no surprise, 
as the population precision is highly sparse with a tre-diagonal structure.
The two graphical precision estimation routines are taking advantage of this.
This thus highlights the importance of employing the graphical nature of the problem in high dimensions.
The shrinkage method performs overall worse than that of Le, but is seen to be more stable, while dominating 
the performance of the sample estimate. 
It also has the benefit of being SPD, and thus has no problems with inversion.
The sample estimate on the other hand performs very poorly when a precision estimate
is required and when $n$ is close to $T$, due to becoming more and more singular.
This is seen in the figure for the Precision matrix, where the norm spikes at $T=100=n$, and then decreases
as the generalised Moore-Penrose inverse takes over the inversion.

% sample size increase
In Figure \ref{fig:fbnorm-vs-sample-size} all estimates are seen to improve with increased sample size,
except for the sample covariance that for a few values of $n$ exhibits the pathological behaviour
close to $n=T$ as described above.
Once again, the \pkg{GraphSPME} algorithm dominates the other three in performance (considering the average from the solid lines).
For small sample sizes, the automatic shrinkage afforded by the method of \citet{touloumis2015nonparametric},
that is implemented in \pkg{GraphSPME},
improves the estimates, as compared to the
un-regularised variants, i.e., the Le method and sample covariance method.
As the sample size is increased, the shrinkage factor is decreased, and necessarily also its effect.
Therefore, the method of \citet{le2022high} is seen to converge to that of \pkg{GraphSPME}, and the sample covariance 
converges to that of the shrinkage method.

% AR-order increase
For the experiment presented in Figure \ref{fig:fbnorm-vs-ar-order}, the sample size is fixed at $n=100$
and the time at $T=40$.
The order of the auto-regressive parameter is increased from 0 
(specifying Gaussian white noise, a fully disconnected graph, and full independence) 
to $p=T-1=39$ for a fully connected graph.
Firstly, note that the sample covariance estimate is stable and does not change much depending on the auto-regressive order.
This is because the sample covariance implicitly always specifies a fully connected graph, and is always maximally flexible.
Secondly, we expect to see the performance of the method of \citet{le2022high} to converge to that of the sample estimate,
while the \pkg{GraphSPME} and \citet{touloumis2015nonparametric} estimate-performance should intertwine more and more.
This behaviour is indeed what can be observed
- the method of \citet{le2022high} benefits early on from the sparse graph specification, but as the graph becomes
fully connected, the method is identical to the implicitly fully connected sample covariance estimate.
Similarly, for the \pkg{GraphSPME} algorithm, it benefits early on from the sparse precision specification,
while for higher values it benefits from the shrinkage afforded by the method of \citet{touloumis2015nonparametric}.
This results in the \pkg{GraphSPME} algorithm dominating all others in performance for all values of the auto-regressive order.

% MAYBE(!?) add table of time-taking thingies 

\section{Discussion} \label{sec:discussion}

This paper describes \pkg{GraphSPME},
a \proglang{C++} header-only library also available in 
\proglang{R} and \proglang{Python} as packages for user friendly applications.
The package takes advantage of and combines the recent methodology estimating
sparse precision matrices with respect to some graphical structure with asymptotic Stein-type shrinkage.
Numerical examples showcase the validity results of this combination.
The package also implements functions for Markov order estimation.
The package can be used for exploratory data analysis, but also as part of 
machine-learning algorithms, advanced statistical regression algorithms or
in the large class of filtering algorithms.

%% -- Bibliography -------------------------------------------------------------
%% - References need to be provided in a .bib BibTeX database.
%% - All references should be made with \cite, \citet, \citep, \citealp etc.
%%   (and never hard-coded). See the FAQ for details.
%% - JSS-specific markup (\proglang, \pkg, \code) should be used in the .bib.
%% - Titles in the .bib should be in title case.
%% - DOIs should be included where available.
\bibliography{references}
%% -- Appendix (if any) --------------------------------------------------------
%% - After the bibliography with page break.
%% - With proper section titles and _not_ just "Appendix".
\newpage
\begin{appendix}
\end{appendix}

\end{document}